\documentclass[5p]{elsarticle}

\usepackage{amssymb}
\usepackage{epsfig}
\usepackage{graphics}
\usepackage{graphicx}
\usepackage{subfigure}
\usepackage{exscale}
\usepackage{latexsym}
\usepackage{makeidx}
\usepackage[section]{placeins}
\usepackage{xtab}
\usepackage{epsf}
\usepackage{bbm}

\usepackage{lineno}

\journal{Nuclear Instruments and Methods in Physics Research Section A}

\begin{document}
\begin{frontmatter}

\title{Neutron spectrometer for fast nuclear reactors}

\author{M.~Osipenko$^a$, 
M.~Ripani$^a$, G.~Ricco$^a$, B.~Caiffi$^b$,
F.~Pompili$^c$, M.~Pillon$^c$, M.~Angelone$^c$,
G.~Verona-Rinati$^d$,
R.~Cardarelli$^e$,
G.~Mila$^f$, S.~Argiro$^f$
}

\address{
$^a$ \it\small INFN, sezione di Genova, 16146 Genova, Italy, \\
$^b$ \it\small Dipartimento di Fisica dell'Universit\`a di Genova, 16146 Genova, Italy, \\
$^c$ \it\small ENEA, Frascati, 00044 Italy. \\
$^d$ \it\small Universit\`a di Tor Vergata, Rome, 00133 Italy. \\
$^e$ \it\small INFN sezione di Roma II, 00133 Italy. \\
$^f$ \it\small Universit\`a di Torino and INFN, Turin, 10125 Italy.
}

\begin{abstract}
In this paper we describe the development and first tests of
a neutron spectrometer designed for high flux environments, such as the ones found in fast nuclear reactors.
The spectrometer is based on the conversion of neutrons impinging on $^6$Li into $\alpha$ and $t$ whose total energy comprises the initial neutron energy and the reaction $Q$-value.
The $^6$LiF layer is sandwiched between two CVD diamond detectors, which measure
the two reaction products in coincidence.
The spectrometer was calibrated at two neutron energies in well known
thermal and 3 MeV neutron fluxes. The measured neutron detection efficiency 
varies from 4.2$\times 10^{-4}$ to 3.5$\times 10^{-8}$ for thermal and 3 MeV neutrons, respectively.
These values are in agreement with Geant4 simulations and close to simple estimates
based on the knowledge of the $^6$Li(n,$\alpha$)$t$ cross section.
The energy resolution of the spectrometer was found to be better than 100 keV
when using 5 m cables between the detector and the preamplifiers.
\end{abstract}


\end{frontmatter}

\section{Introduction}\label{sec:intro}
The standard nuclear reactor on-line diagnosis is performed with fission chambers.
Solid state detectors have a number of features which distinguish them from gas filled counters.
In particular, much faster carrier mobility allows for a rapid charge collection and therefore high counting rate. The high density of the active volume permits spectroscopic
measurements with compact detectors which do not alter local conditions.
The sensitivity to $\gamma$s is also lower for solid state detectors because of lower $Z$
and smaller size. However, a limitation in the application of solid state detectors to nuclear diagnostics stems from the fact that large non-ionizing energy deposition damages the crystalline structure and alter the properties of the detector. This leads to an increase of leakage current and reduction
of charge collection efficiency. Silicon detectors are strongly affected by this type of damage
and cannot be used in high radiation environment. Thanks to higher displacement energy (43 eV) 
and lower $Z$ value, diamond detectors feature an order of magnitude larger
resistance to non-ionizing doses~\cite{cvd_rad_hard}.
Therefore, several studies are available in literature that present their application as diagnostic tools in nuclear facilities.

In addition to radiation hardness diamond detectors exhibit also
a low intrinsic noise at high temperatures thanks to their large bandgap of 5.5 eV.
This permits to use diamond detectors in fission reactors where operating
temperature may reach 700 degrees and near fusion plasma chamber.
However, presence of impurities in the diamond crystal creates levels inside the bandgap
increasing leakage current and therefore detector noise. The impurity density limited
usage natural diamonds as radiation detectors.
The advent of the CVD diamond growing technique enhanced the quality of diamond detectors and allowed practical applications.

Nuclear reactors in which the neutron spectra are dominated by energies well above thermal energies are called fast reactors.
These include reactors featuring heavy or low density moderators. A more energetic neutron spectrum
allows for a lower production rate of radioactive waste and for the burn-out of a fraction
of the long lived actinides. However, both the reactor dynamics
and burn-out of fuel and actinides depend on the neutron spectrum. Conventionally the neutron spectrum
in a reactor is measured by activation foil analysis.
Indeed, the activation of an isotope can be related to the convolution of the neutron flux
with the isotope activation cross section.
But this complex, off-line, procedure
introduces large systematic uncertainties. A simple on-line technique is necessary for characterization
of reactor transients. For this purpose we developed a novel neutron spectrometer based on a $^6$Li
converter sandwiched between two CVD diamond detectors. The energy of the incident neutron
converts completely into the energy of charged particles through the $^6$Li(n,$\alpha$)$t$ reaction.
This allows for event-by-event neutron energy measurement with the advantages of a solid state detector.
Moreover, the $^6$Li(n,$\alpha$)$t$ reaction is highly exothermic with $Q=4.7$ MeV, which
permits to reduce background by imposing a high detection threshold. The coincidence
between two crystals suppresses further noise and competing reactions.

In this article we describe the spectrometer assembly and two calibration
experiments. The details of detector development are given in section~\ref{sec:det}.
Measurements of the detector response to thermal neutrons, made at a TRIGA reactor, are discussed in section~\ref{sec:triga},
while the fast neutron experiment at a DD-fusion source is detailed in section~\ref{sec:fng}.

\section{Detector Construction}\label{sec:det}
The sandwich spectrometer prototype was built from two Single-Crystal Diamond detectors (SCD),
identified as SCD398 and SCD1517, grown at the laboratories of the University of Rome ``Tor Vergata''.

\begin{figure}[h]
\begin{center}
\includegraphics[bb=1cm 0cm 20cm 26cm, scale=0.3, angle=270]{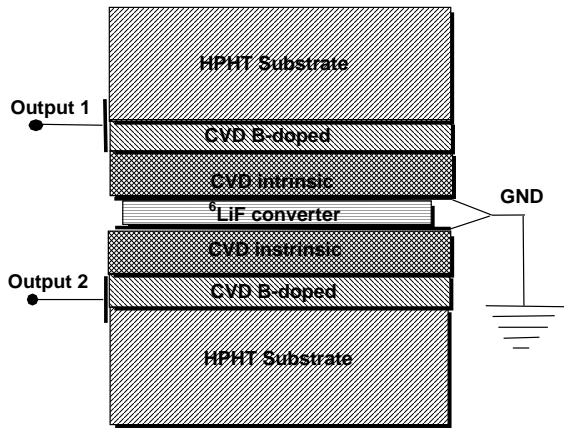}
\caption{\label{fig:sdw_geom}Design of the sandwich spectrometer with its main components:
HPHT diamond substrate, B-doped diamond layer, intrinsic diamond layer, metallic contact
and LiF converter.}
\end{center}
\end{figure}

Each SCD had been grown on a $4\times 4\times 0.4$ mm$^3$ HPHT substrate, and had the same structure shown in Fig.~\ref{fig:sdw_geom}. These detectors are composed of three main layers: p-type diamond, intrinsic diamond and metal contact. This diode-like structure is described in Refs.~\cite{fulvio_r1,fulvio_r2}
and allows the readout of signals from the intrinsic diamond layer without removal of the HPHT substrate.
The degenerate p-type layer acts as an ohmic contact.
Instead, the anode, creates a Schottky junction with the underlying intrinsic diamond.
In this configuration the electric current generated by the passage of an ionizing particle
in the depletion layer flows across the detector without any barrier. In fact, the Schottky
junction at the intrinsic diamond-metal interface accelerates electrons leaving the diamond bulk.

\begin{figure}[h]
\begin{center}
\includegraphics[bb=1cm 0cm 20cm 26cm, scale=0.3, angle=270]{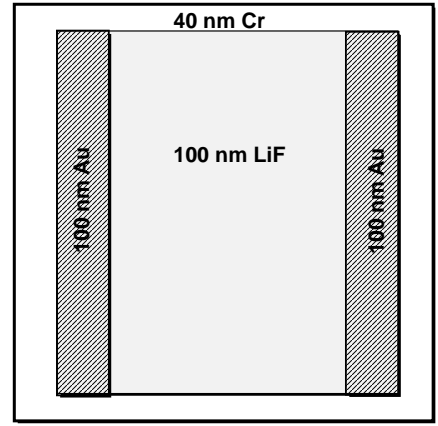}
\caption{\label{fig:sdw_deposit}Gold strips and LiF converter
with Chromium contact underneath, 
deposited on top of the intrinsic CVD diamond crystal.}
\end{center}
\end{figure}

A 3 mm $\times$ 3 mm $\times$ 40 nm Chromium layer had been deposited as the top contact;
it was also used as a sticking layer for the two narrow (0.4 mm $\times$ 3 mm $\times$ 80 nm) gold strips
shown in Fig.~\ref{fig:sdw_deposit}. These strips were used to readout anode signals.
On top of the Chromium layer in between the Gold strips a further layer
of 3 mm $\times$ 2.2 mm x 100 nm of $^6$LiF was then deposited on the contact.
The LiF compound was chosen as a neutron converter because of its chemical neutrality.
The selected LiF was enriched with $^6$Li isotope to 96\%.
The LiF layer is deposited on the top of metallic anode in such a way that $\alpha$ and $t$
can easily penetrate into the intrinsic diamond depletion layer with minimal loss of energy.
In fact the assembly leaves only 50 $\mu$m gap for the central ground microwire.

The active volume of each detector consists of its intrinsic diamond layer.
The thicknesses for the Boron doped (P-type) and intrinsic diamond layers were different for the two detectors,
as shown in Table~\ref{tab:scd_character}. The thickness of the chromium contact, and of the $^6$LiF layer,
instead, were the same within the uncertainties of the measurement system of the thermal evaporator.

The ``sandwich'' structure is realized by means of two small double-layers Printed Circuit Boards (PCB)
(35 $\mu$m golden copper on 0.8 mm FR4), to which diamonds are glued. Each PCB has a ground plane on one side
and signal traces on the other. The P-type layers are connected to the signal traces with a droplet of conductive glue.
The top electrode, which is shared by the two diamonds, is connected by two microwires running along
the Gold strips, not covered by LiF, and soldered to the ground plane through four vias.
The wires are made of Copper (diameter 50 $\mu$m) plated with 2 $\mu$m of galvanically deposited gold.
Each PCB has four golden-plated connectors which guarantee the integrity of the ground plane and the routing of electrical signals,
but also a precise mechanical alignment between the diamonds and the wires when the ``sandwich'' is closed,
i.e. when the two PCBs are faced one to each other. The stability of the electrical contact between the wires
and the Gold strips of both diamonds is ensured by a U-shaped spring that ``closes'' the sandwich
by operating a small force pushing the PCBs one toward the other. The two PCBs have different length,
allowing RG62 signal cables to be soldered on dedicated pads on the longer board.

\begin{table}[!h]
\begin{center}\label{tab:scd_character}
\caption{Diamond detector characteristics.} \vspace{2mm}
\begin{tabular}{|c|c|c|c|c|} \hline
  SCD   & P-doped  & Intrinsic & Cr      & LiF   \\ 
        & layer    &  layer    & contact & layer  \\ 
        & thick.   & thick.    & thick.  & thick. \\ 
        & [$\mu$m] & [$\mu$m]  & [nm]    & [nm]   \\ \hline
SCD398  & 23       & 22        & 40      & 100    \\ \hline
SCD1517 & 15       & 49        & 40      & 100    \\ \hline
\end{tabular}
\end{center}
\end{table}

\section{Measurement at TRIGA reactor}\label{sec:triga}
In order to characterize the detector and to evaluate the absolute amount
of $^6$Li isotope deposited on the sensitive detector area a measurement in a calibrated
thermal neutron flux was performed. The measurement was carried out at LENA of Pavia University
on the TRIGA reactor. The reactor has variable power from few kW up to 250 kW, accurately
monitored by an in-core compensated fission chamber system~\cite{triga_mon}. It features several irradiation channels,
among which a large graphite thermal column. The neutron flux in the column is almost
perfectly Maxwellian reaching values of 1.2$\times 10^{10}$ n/cm$^2$/s at 250 kW.
The detector was installed in the location inside the thermal column where the neutron flux
was carefully determined in Ref.~\cite{triga_flux}.
The data were acquired at different level of reactor power from 1 kW up to 100 kW.

\subsection{Experiment}
The sandwich spectrometer prototype under test was made of two CVD diamonds
SCD398, identified as channel 1, and SCD1517, identified as channel 2.
The signals from the two diamonds were connected via 2.5 m RG62 cables to Silena 205
charge preamplifiers. The amplified signals were driven to the Ortec 673 Spectroscopy amplifiers
where they were further amplified and filtered with 2 $\mu$s shaping time.
The final signals were acquired by CAEN V785 peak sensing ADC.
The trigger signals were taken from CAEN V895 discriminator, which was connected to
the secondary preamplifier output, amplified by Ortec 460
delay line amplifiers with 200 ns shaping time.

The Data AcQuisition (DAQ) system was based on VME modular electronics. In particular,
Concurrent Tech. VX813-09x single board computer was used as VME controller as well as
the acquisition host. The VX813-09x run 32-bit Centos 6 Linux operating system with
native Tsi148 VME controller drivers. The CAEN V785 peak sensing ADC was configured
to generate interrupts on VME bus when the number of events in its multi-event buffer exceeded the imposed threshold.
When the interrupt was received by the VX813/09x board it copied all the data
from the V785 buffer through a fast DMA MBLT transfer and sent them together with interrupt timestamp to a secondary thread.
This secondary thread was buffering and saving the data on a fast SATA SSD. The amount of data
and event rate were relatively small, compared to the host memory size,
thus VME transfer and disk writing speed were sufficient
to run acquisition without additional delays. For the measurement of the system dead time
the discriminated signals were counted on CAEN V830 scaler during runtime.

\subsection{Data analysis}
\begin{figure}[!h]
\begin{center}
\includegraphics[bb=4cm 0cm 20cm 26cm, scale=0.35, angle=270]{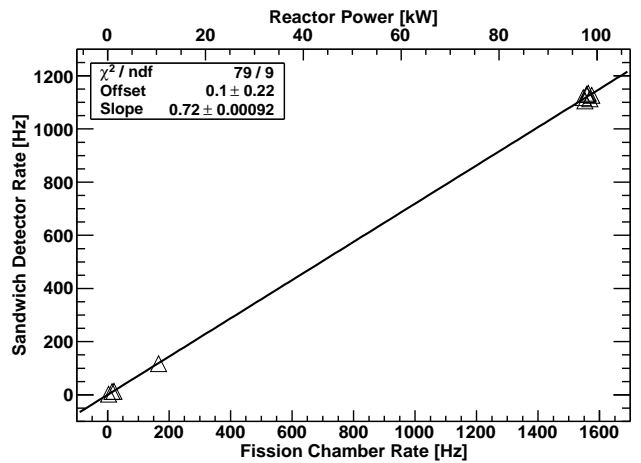}
\caption{\label{fig:pavia14_fc_swd_rates} Correlation between sandwich detector event rate,
fission chamber rate and reactor power.}
\end{center}
\end{figure}

For the absolute normalization of the measurement two independent informations were used.
The first was provided by the TRIGA power monitoring system combined with the
power-to-flux conversion from Ref.~\cite{triga_flux}. For the second an independent
fission chamber with known sensitivity of $3\times 10^6$ n/cm$^2$s/cps~\cite{altieri_fc}
was installed. The fission chamber signal was amplified by Ortec 142B charge sensitive preamplifier and
Ortec 572A Spectroscopy amplifier and then discriminated by the Ortec 934 discriminator.
The discriminator logical output was connected to the CAEN V830 scaler, counting
all the signals within DAQ runtime.
The measured fission chamber signal rate was found to be in good agreement with
the TRIGA power monitoring system. The correlation between sandwich detector event rate
and fission chamber signal rate was found to be linear within uncertainties
on the interval of two orders of magnitude, as shown in Fig.~\ref{fig:pavia14_fc_swd_rates}.
The locations of the fission chamber and sandwich spectrometer were different
leading to somewhat large $\chi^2$ value.

%

Since the DAQ system was configured to acquire both ADCs when a trigger in any of the two
channels occurred (OR configuration), the coincidence events were selected off-line.
The selection was based on the application of an off-line
threshold to both ADCs, determined as the ADC channel above which the measured yield was more than 100 times higher
than a possible tail from the ADC pedestal. The discarded events were mainly those in which one
of two reaction products $\alpha$ or $t$ was lost. This occurred when one of reaction products
was emitted either in the 50 $\mu$m gap between the two crystals or into a non-sensitive
area of the crystal.

\begin{figure}[!ht]
\begin{center}
\includegraphics[bb=3.8cm 0cm 21.5cm 26cm, scale=0.3, angle=270]{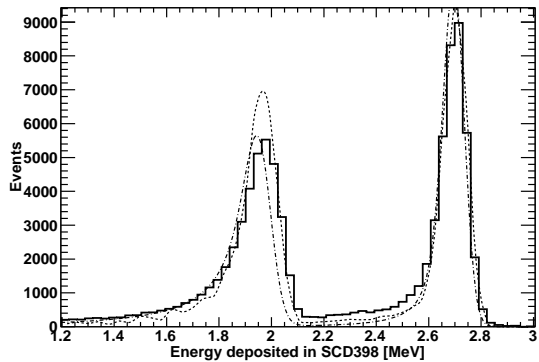}
\caption{\label{fig:edep_i}Energy deposited in the single diamond of the spectrometer
in comparison with Geant4 (dashed line) and MCNP (dot-dashed line) simulations.
In MCNP simulations 100 nm Cr contacts were used, while in Geant4 they were set to 40 nm.}
\end{center}
\end{figure}

The calibration of the energy deposited in each crystal was based on the energy
of the produced $t$, corrected for the energy lost in LiF layer and Cr contacts.
The ADC pedestal peak position was taken as the zero energy point.
The obtained deposited energy distribution in the SCD398 crystal is shown in Fig.~\ref{fig:edep_i}.
The high energy peak, due to the absorption of $t$, has almost Gaussian shape.
The resolution of $t$-peak was found to be $42$ keV for SCD398 and $22$ keV for SCD1517.
These values are only slightly larger than the measured pedestal widths of $38.5$ and $18.5$ keV, respectively.
The intrinsic resolution due to the energy loss in LiF layer and Cr contacts is estimated
to be about $5$ keV.

The lower energy peak in Fig.~\ref{fig:edep_i} is due to the absorption of $\alpha$. It exhibits
a significant low-energy tail because of the higher energy loss of $\alpha$s in LiF and Cr.

\begin{figure}[!h]
\begin{center}
\includegraphics[bb=4.5cm 0cm 20cm 26cm, scale=0.35, angle=270]{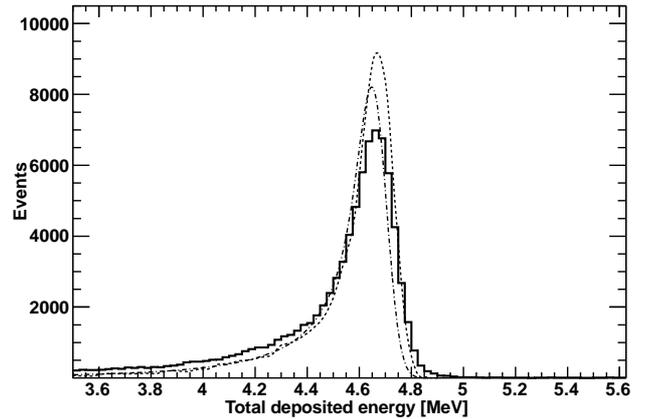}
\caption{\label{fig:triga_edep_tot}The same as in Fig.~\ref{fig:edep_i}, but for
the total energy deposited in the spectrometer.}
\end{center}
\end{figure}

The sum of the two energies deposited in the two crystals gives the incident neutron energy
increased by the reaction $Q$-value. For thermal neutrons the incident energy is much lower than
the detector energy resolution and therefore in this case the total energy must be equal
to the reaction $Q$-value (4.7 MeV). Summing up the energies of the two crystals indeed leads
to an asymmetric peak at around 4.7 MeV, as shown in Fig.~\ref{fig:triga_edep_tot}.
The peak resolution was found to be $73$ keV,
compared to the intrinsic resolution of $21$ keV due to energy losses in LiF layer and Cr contacts.

%

The major goal of this measurement was the evaluation of absolute efficiency of the spectrometer
to detect thermal neutrons. The spectrometer inefficiency to the produced $t$+$\alpha$ pair was fairly low
due to small angular acceptance of the 50 $\mu$m gap between the crystals and almost complete
coverage of LiF layer by the underlying Cr anode.
However, the amount of $^6$Li deposited by evaporation on the diamond Cr contacts was known with a large uncertainty.
To determine it precisely a comparison between the data and Monte Carlo was used.

The sandwich detector response to the neutron flux was modeled using Geant version 4.9.5.p01.
The detector was described in details,
including:
PCB support with its metalization, HPHT diamond substrate, B-doped CVD diamond cathode substrate, intrinsic CVD diamond,
Cr anode layer, LiF layer, Gold strips and Gold plated Cooper wires.

The neutron flux of the TRIGA thermal column~\cite{triga_flux} was generated isotropically on the spherical
surface around the spectrometer of area about 1 cm$^2$. The events with the energy deposited
in any of two crystals of the sandwich sensitive volume above the threshold were selected.

In order to normalize the simulations to the same neutron flux seen by the spectrometer
the reconstructed Geant4 distributions were rescaled by the factor:
\begin{equation}
L_{sim}= \phi_n^{tot} t_{run} \frac{S_{gen}}{N_{gen}} ~,
\end{equation}
\noindent where $\phi_n^{tot}$ is the total neutron flux measured by fission chamber,
$t_{run}$ is DAQ runtime, $N_{gen}$ is the total number of neutrons generated on the surface of area $S_{gen}$.
The thickness of LiF was first assumed to be equal to the nominal value of 200 nm
and then adjusted as a free parameter to reproduce the data. The obtained in this way
thickness of LiF was found to be 80 nm. This effective number includes also all other
spectrometer inefficiencies not accounted for in Geant4 simulation.

In order to compare Geant4 simulations to the data, an additional Gaussian smearing due to
electronic readout noise was added to the reconstructed energies.
The obtained simulations describe the data fairly well, in particular in the region of $t$-peak,
which has almost a Gaussian shape. Instead the $\alpha$-peak is more asymmetric and features a larger
l.h.s. tail due to higher energy loss of $\alpha$ particles in LiF and Cr.
The total deposited energy distribution is also well reproduced.

Similar simulations were performed with MCNP, but with Cr contacts of 100 nm and 200 nm LiF thickness
which determined a larger energy loss of $\alpha$ and slight shift in its peak position.

\begin{figure}[!h]
\begin{center}
\includegraphics[bb=4.5cm 0cm 20cm 26cm, scale=0.35, angle=270]{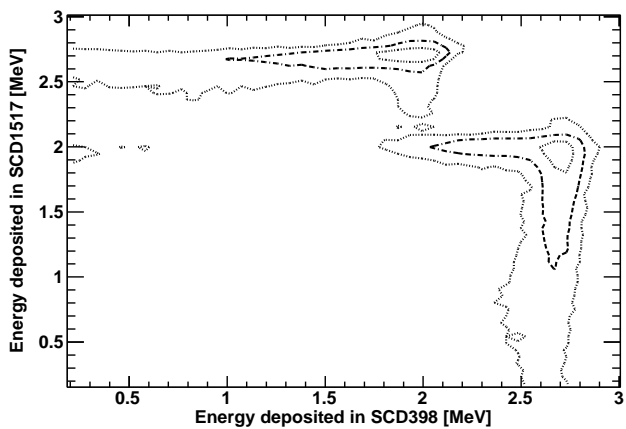}
\caption{\label{fig:triga_edep_cor}Correlation between deposited energy spectra in the two diamonds of the spectrometer.
$t$-peak at 2.7 MeV one crystal corresponds to $\alpha$ peak at 2 MeV in the other.
The long tail of the $t$-peak in SCD398 is due to energy leak out of crystal depletion layer.}
\end{center}
\end{figure}

The correlation between the signals in the two crystals, shown in Fig.~\ref{fig:triga_edep_cor}, revealed
an important difference: the enhanced tail following $t$-peak in SCD398.
This was related to the crystal depletion layer thickness, in case of SCD398 smaller than the 2.7 MeV $t$ range in the diamond.
In fact, those $t$ incident on SCD398 almost perpendicular to its surface were not completely absorbed.
This behavior was reproduced by Geant4 simulations using the expected depletion layer thickness of 18 $\mu$m.

The fast charge amplifier developed by R.~Cardarelli~\cite{cardarelli_amp} was also used for readout together
with SIS3305 5 Gs/s digitizer. It showed results consistent with the standard charge sensitive electronics
reducing drastically dead time. However, the energy resolution was found to be at least two times worse.

\section{Measurement at FNG}\label{sec:fng}
The response of the spectrometer to fast neutrons was measured at FNG facility of ENEA~\cite{fng_facility}.
FNG is a neutron generator based on the T(d,n)$\alpha$ and D(d,n)$t$ fusion reactions to produce a nearly
isotropic source of $5\times 10^{10}$ 14-MeV n/s or $5\times 10^8$ n/s 2.5-MeV n/s, respectively.
The neutron source strength is continuously monitored by counting the recoiling $\alpha$ or $t$ particles
associated with each neutron produced by the DT or DD reaction. The recoiling charged particles are counted
by means of a small silicon surface barrier detector incorporated in the beamline.
The resulting uncertainty on the source strength measurement is lower than 4\%.
The FNG target is installed at 4 m above the floor and 4 m from walls and ceiling thus reducing
the contribution from neutron rescattering on environmental materials.

\subsection{Experiment}
\begin{figure}[!ht]
\begin{center}
\includegraphics[bb=4.5cm 1.5cm 20cm 27.5cm, scale=0.35, angle=270]{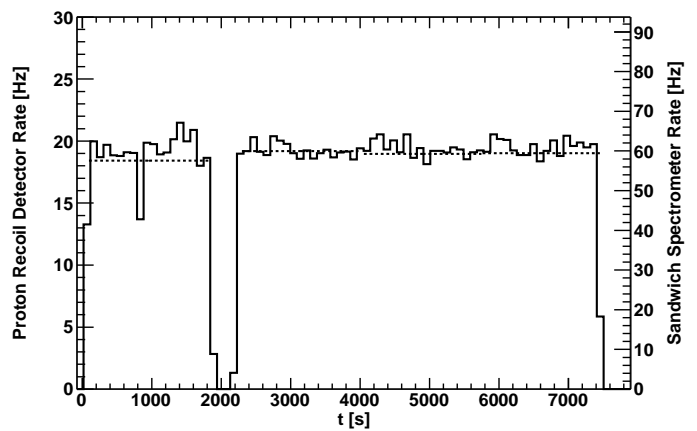}
\caption{\label{fig:fng_p_recoil_swd_rates}Rate of the FNG recoil detector (histogram)
in comparison to the sandwich spectrometer rate (dashed lines) rescaled for comparison.}
\end{center}
\end{figure}

The detector was installed at 2.5 cm in front of the TiT target (zero degrees with respect to $D$ beam).
During the experiment FNG was operated in DD-mode with the beam current of 1 mA and the total
neutron yield of $5.7\times 10^8$ n/s.
The total neutron yield was continuously monitored by the recoil detector,
whose rate variations during the experiment is shown in Fig.~\ref{fig:fng_p_recoil_swd_rates}.
Moreover, to better determine the flux of fast neutrons at the spectrometer location,
two $^{115}$In foils were installed in front and at the back of the spectrometer.
In order to calibrate the ADC energy scale a plastic moderator was installed behind the detector.

DAQ was run in coincidence (AND) configuration at the total rate of about 60 Hz. At such a low event rate
DAQ dead time was negligible.

\subsection{Data analysis}
The irradiated $^{115}$In 95.7\% NA foils were measured at 60\% relative efficiency HPG detector immediately after
the experiment and the strengths of a few $\gamma$ emission lines of two reactions were determined. The first reaction
was $^{115}$In(n,n$^\prime$)$^{115m}$In, where the produced excited state decays with 4.49 hours half-life
emitting 336 keV $\gamma$ with 45.8\% probability. This reaction is mostly sensitive to the 2.5-3 MeV
neutrons from DD-fusion. The second reaction measured by HPG was $^{115}$In(n,$\gamma$)$^{116m}$In
where the produced metastable nucleus decayed with 54.15 minutes half-life emitting a number of $\gamma$s with various
probabilities: 417 keV (29.2\%), 819 keV (11.5\%), 1097 keV (56.2\%), 1294 keV (84.4\%), 1507 keV (10\%).
The two measured activities together with detailed neutron spectrum calculated by MCNP simulations
were used by SAND II~\cite{sand2} unfolding procedure to obtain the absolute neutron flux as a function of neutron energy.
The MCNP simulations described in details the FNG neutron source and were carefully
tested in Ref.~\cite{fng_mcnp}.

Because of very low signal rate in this measurement the ADC spectra were dominated by accidental coincidences.
These were mostly events due to electronic noise and the elastic neutron scattering on carbon nuclei.
The events in which both ADCs were above the channel where the number of counts
was more than 100 times larger than the pedestal tail were selected as coincidences.

The ADC energy calibration was accomplished using the position of the thermal neutron
induced $t$-peak, clearly visible in both diamonds.
The zero energy scale was obtained from the pedestal peak position.

Preliminary simulations showed that most of the 3 MeV neutron events deposit
more than 1.4 MeV in each crystal. Hence, the off-line threshold of 1.4 MeV was applied
in order to remove the elastic scattering contribution and other accidental coincidences.

\begin{figure}[!h]
\begin{center}
\includegraphics[bb=4.5cm 0cm 20cm 26cm, scale=0.35, angle=270]{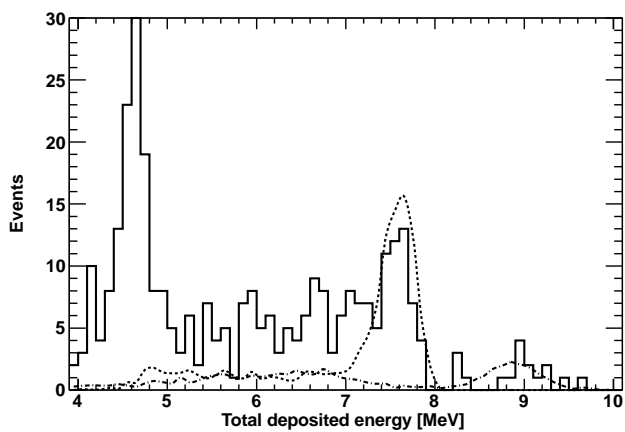}
\caption{\label{fig:fng_etot_swd3}Total energy deposited in the spectrometer
in comparison with Geant4 simulations of DD neutrons (dashed line) and
DT neutrons (dot-dashed line). The peak at 4.7 MeV is due to thermalized neutrons
used for energy calibrations, the tail at the left of 7.7 MeV peak (3 MeV neutrons)
is partially due to DT neutron inelastic reactions in the diamond and
due to the tail of scattered 3 MeV neutrons.}
\end{center}
\end{figure}

Once the accidental coincidences were removed, the energies deposited in the
two crystals were summed up in the total deposited energy showed in Fig.~\ref{fig:fng_etot_swd3}.
The obtained distribution exhibits a large peak at 4.7 MeV due to thermal neutron
detection as well as lower peak at 7.7 MeV due to direct 3 MeV neutrons from DD source.

Comparisons of data to Geant4 simulations performed with a realistic neutron spectrum
revealed the complex structure of the measured distribution.
Indeed, few events seen at 9 MeV were due to direct 15 MeV neutron conversion
into $^9$Be$+\alpha$, where $\alpha$, produced in the first diamond, traveled into the second crystal
to generate the coincidence.

The l.h.s. shoulder of the 7.7 MeV peak is made of two contributions of similar magnitude:
15 MeV neutron conversion in 3$\alpha$ and the tail of DD peak from initial neutron spectrum.
This tail was produced by the neutron rescattering from surrounding materials.

The resolution of the detector in this experiment was slightly worse than in the TRIGA
measurement as one can see from the thermal peak $\sigma$ of $\sim 90$ keV.
This was related to a higher level of environmental electronic noise in the experimental hall.

\begin{figure}[!h]
\begin{center}
\includegraphics[bb=4.5cm 0cm 20cm 26cm, scale=0.35, angle=270]{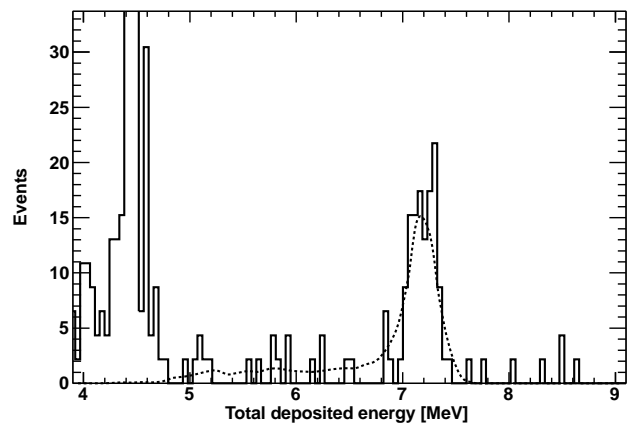}
\caption{\label{fig:fng_etot_swd1_old} The same as if Fig.~\ref{fig:fng_etot_swd3} but measured
at 90$^\circ$ with respect to deuteron beam with Prototype II. Peak width is almost factor of two smaller than at 0$^\circ$.}
\end{center}
\end{figure}

Prototype II was installed at 90 degrees with respect to $D$ beamline
at 3 cm distance from the target center. The same analysis was performed on its data.
The obtained results (see Fig.~\ref{fig:fng_etot_swd1_old}) are in good
agreement with Geant4 simulations.
The expected neutron spectrum features a peak at 2.45 MeV with twice smaller width (77 keV instead of 177 keV at 0 degrees). In the measured data the peak widths are enhanced by the spectrometer energy resolution,
but the ratio between 90 and 0 degrees widths is similar: 114 keV and 204 keV, respectively.

\section{Backgrounds}\label{sec:bkg}
Thanks to highly exothermic reactions on $^6$Li backgrounds can be neglected
when the maximum neutron energy (with significant flux) remains below 6 MeV.
In fact, the most probable reaction, neutron elastic scattering off carbon nuclei,
deposits only $\sim$25\% of its energy in the detector. Therefore it can be easily removed
by a high threshold (1.4 MeV for neutron up to 6 MeV) with minimal efficiency loss.
All the inelastic reactions on carbon, except for unlikely radiative capture and excitation
of the first $^{12}$C level, occur above 6 MeV.

The spectrometer has a low sensitivity to $\gamma$-rays, thanks to its low $Z=6$ value.
Moreover, due to the small thickness of the crystals (50 $\mu$m) a significant energy loss
by electrons in the spectrometer sensitive volume is very unlikely.

Therefore, we neglected all backgrounds except for contributions from other parts
of the neutron spectrum. These other parts of the neutron spectrum seen in FNG experiment
are mostly due to T contamination in the TiD target, yielding 15 MeV neutrons,
and due to the tail of 3 MeV neutrons rescattered from surrounding materials.

\section{Systematic uncertainties}\label{sec:sys_err}
The low statistics of FNG measurements allowed only for 10\% statistical precision.
Thus, we considered only major contributions which were of the order of
these uncertainties.

The largest uncertainties come from the calibrating neutron flux normalizations.
In TRIGA measurement the fission chamber sensitivity of $3\times 10^6$ n/cm$^2$s/cps~\cite{altieri_fc}
was established with uncertainty of 8\%,
while the neutron flux in the TRIGA thermal column as a function of reactor power had 2.5\% precision.

In FNG measurement we relied on the $^{115}$In activation foil analysis which
had an overall uncertainty of 20\%. This value combines both uncertainties
due to the position mismatch and due to the spectrum reconstruction from the excited level activity measurement.

\section{Results}
\begin{figure}[!ht]
\begin{center}
\includegraphics[bb=4.5cm 0cm 20cm 26cm, scale=0.35, angle=270]{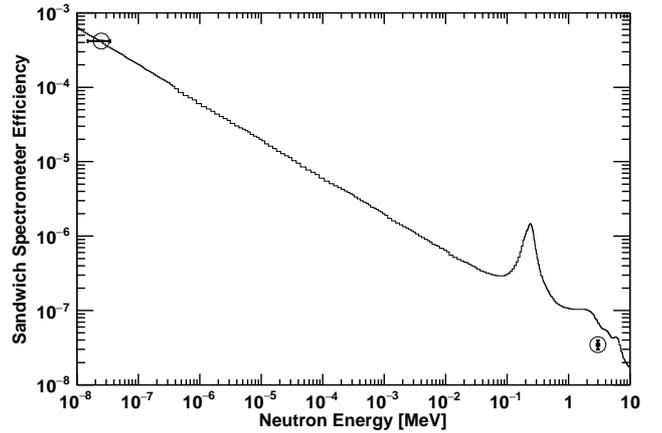}
\caption{\label{fig:eff_swd}Neutron detection efficiency of sandwich spectrometer
for the neutrons incident on 0.066 cm$^2$ active surface
in comparison with $^6$Li(n,t)$\alpha$ cross section rescaled by surface $^6$Li atom density.}
\end{center}
\end{figure}

The detection efficiency of the sandwich spectrometer prototype for neutrons incident
on its $^6$LiF layer area of 0.066 cm$^2$  was measured at two
neutron energies: thermal (25 meV) and 3 MeV. The obtained energy dependence was found to be
in good agreement with the known $^6$Li(n,$\alpha$)t cross section rescaled by the surface atom density
as shown in Fig.~\ref{fig:eff_swd}.
At 3 MeV the measured efficiency was 30\% lower than the cross section,
but well described by Geant4 simulations. This is related to a broader energy distribution
of reaction products, some of which carry energy below the selected threshold.

In order to obtain detector sensitivity in cps/(n/cm$^2$s) the efficiency has to
be multiplied by the spectrometer active surface area of 0.066 cm$^2$.

\begin{table}[!h]
\begin{center}\label{tab:delta_r}
\caption{Measured efficiency of the neutron spectrometer
for the neutrons incident on 0.066 cm$^2$ active surface.} \vspace{2mm}
\begin{tabular}{|c|c|} \hline
  $E_n$   & Efficiency \\ \hline
 25 meV   & 4.2$\times 10^{-4}$ $\pm$ 0.00003\%$_{stat.}$ $\pm$ 8\%$_{sys.}$  \\ \hline
 3 MeV    & 3.5$\times 10^{-8}$ $\pm$ 14\%$_{stat.}$ $\pm$ 20\%$_{sys.}$ \\ \hline
\end{tabular}
\end{center}
\end{table}

\section{Conclusions}\label{sec:conclusions}
A neutron spectrometer for fast nuclear reactors based on $^6$LiF converter sandwiched
between two CVD diamond detectors was developed. This work was mostly dedicated
to the measurements of the spectrometer efficiency as a function of neutron energy.
Because of small efficiency values only high neutron flux facilities could be used
for this purpose.
The first prototype was assembled
and calibrated to thermal neutrons at TRIGA reactor of LENA and to 3 MeV neutrons at FNG facility of ENEA.
The energy dependence of the efficiency of the spectrometer was found to be in good agreement with expectations
based on Geant4 and MCNP simulations.
The neutron energy resolution was found
to be as low as 73 keV (RMS).
The most important contribution in the resolution function was given by electronics noise,
while the intrinsic resolution of the spectrometer was estimated to be 21 keV.
The latter derives
from the stochastic energy loss in LiF converter and Cr contacts.

Due to its low efficiency the spectrometer is suited for neutron fluxes $>10^7$ n/cm$^2$s.
Moreover, because of a higher sensitivity to thermal neutrons and opening
of inelastic reaction channels on $^{12}$C above 6 MeV the most appropriate
usage of the spectrometer is the measurement of fission spectrum.
Such a neutron spectrum can be found in fast fission reactors
or spent fuel elements.

\section*{Acknowledgement}
Authors would like to acknowledge the excellent support provided during the experiments
by the staff and technical services of LENA and FNG facilities.
We thank Prof.~Saverio Altieri of Pavia University for the help with fission chamber measurements in TRIGA reactor.
This work was supported by the Istituto Nazionale di Fisica Nucleare INFN-E project.

\bibliographystyle{elsarticle-num}
\bibliography{sdw_calib}

\begin{thebibliography}{10}
\expandafter\ifx\csname url\endcsname\relax
  \def\url#1{\texttt{#1}}\fi
\expandafter\ifx\csname urlprefix\endcsname\relax\def\urlprefix{URL }\fi
\expandafter\ifx\csname href\endcsname\relax
  \def\href#1#2{#2} \def\path#1{#1}\fi

\bibitem{cvd_rad_hard}
W.~de~Boer, et~al., Phys. Status Solidi 204 (2007) 3009.

\bibitem{fulvio_r1}
M.~Angelone, et~al., Neutron detectors based upon artificial single crystal
  diamond, Vol.~56, IEEE TRANSACTIONS ON NUCLEAR SCIENCE, 2009, p. 2275.

\bibitem{fulvio_r2}
S.~Almaviva, et~al., Thermal neutron dosimeter by synthetic single crystal
  diamond devices, Vol.~67, Applied Radiation and Isotopes, 2009, p. 183.

\bibitem{triga_mon}
A.~{Borio~di~Tigliole}, et~al., Home-made refurbishment of the instrumentation
  and control system of the triga reactor of the university of pavia, 2008.

\bibitem{triga_flux}
N.~Protti, S.~Bortolussi, M.~Prata, P.~Bruschi, S.~Altieri, D.~Nigg, Neutron
  spectrometry for the university of pavia triga thermal neutron source
  facility, Vol. 107, TRANSACTIONS OF THE AMERICAN NUCLEAR SOCIETY, 2012, p.
  1269.

\bibitem{altieri_fc}
A.~Altieri, Private communication.

\bibitem{cardarelli_amp}
R.~Cardarelli, A.~D. Ciaccio, L.~Paolozzi, Nucl. Instr. and Meth. A 745 (2014)
  82.

\bibitem{fng_facility}
M.~Martone, M.~Angelone, M.~Pillon, J. of Nucl. Mater. B 212 (1994) 1661.

\bibitem{sand2}
W.~McElroy, S.~Berg, T.~Crockett, R.~Hawkins, A computer-automated iterative
  method for neutron flux spectra determination, report AFWL-TR 67-41 (1967).

\bibitem{fng_mcnp}
M.~Angelone, M.~Pillon, P.~Batistoni, M.~Martini, M.~Martone, V.~Rado, Rev. of
  Sci. Instrum. 67 (1996) 2189.

\end{thebibliography}

\end{document}